\documentclass[epj]{svjour}

\usepackage{graphicx}
\usepackage{fancyhdr}
\def\makeheadbox{{
 }}
\def\mytitle{My title} 
\def\myauthors{My name}  
\def\mytype{My type of session}
\def\mysession{My session}

\def\mytitle{MSSM Higgs-bosons from bottom quarks} 
\def\myauthors{Alexander M\"uck}    
\def\mytype{Contributed Talk}    
\def\mysession{Colliders - Higgs Phenomenology}

\begin{document}
\title{Precise predictions for MSSM Higgs-boson production 
in bottom-quark fusion}
\author{Alexander M\"uck\inst{1, 2}
\thanks{\emph{Email:} alexander.mueck@psi.ch}%
}                     
%
%
\institute{
Institut f\"ur Theoretische Physik E, RWTH Aachen, 
D-52056 Aachen, Germany
\and 
Paul Scherrer Institut, W\"urenlingen und Villigen, 
Ch-5232 Villigen PSI, Switzerland
}
%
\date{}
\abstract{
The main production mechanism for supersymmetric Higgs particles 
at hadron colliders crucially depends on the size of their 
Yukawa couplings to bottom quarks. For sufficiently large $\tan\beta$ 
the total cross section for some of the neutral Higgs bosons 
in the MSSM is dominated by bottom-quark fusion. After an
introduction to bottom-associated Higgs production, we discuss the 
complete ${\cal O}(\alpha)$ electroweak and 
${\cal O}(\alpha_{\mathrm{s}})$ strong corrections for the 
$\mathrm{b}\bar\mathrm{b}$-fusion channel in the MSSM.
Choosing proper renormalization and input-parameter schemes, 
an improved Born approximation, constructed from previously known 
results, can absorb the bulk of the corrections so that 
the remaining non-universal corrections are typically of the 
order of a few per cent. Numerical results are discussed for the 
SPS benchmark scenarios.
} 
\maketitle
\section{Introduction}
\label{intro}
The Higgs mechanism is a cornerstone of the Standard Model (SM) and
its supersymmetric extensions. Thus, Higgs bosons are intensively
searched for at the upgraded proton--antiproton collider Tevatron,
followed in the near future by the  proton--proton collider LHC. In
this talk, we concentrate on the  precise prediction of the total
Higgs-boson  production cross section at the LHC based on the
results  in~\mbox{Ref.~\cite{Dittmaier:2006cz}}. 

In the SM, the total production cross section for Higgs bosons 
$\mathrm{H}$ at the LHC is dominated by gluon fusion. Higgs
radiation off  bottom quarks~\cite{Raitio:1978pt}
\begin{equation}
\mathrm{p}\bar{\mathrm{p}} / \mathrm{p}\mathrm{p} \to \mathrm{b}\bar{\mathrm{b}}\,\phi^0 \!+\!X 
\label{eq:procs_hadron}
\end{equation}
with $\phi^0 = \mathrm{H}$,  is a negligible contribution. The
relevant bottom Yu\-ka\-wa coupling  $\lambda^{\rm SM}_\mathrm{b}$
is known to be small because it is determined by the ratio  of the
small bottom-quark mass $m_\mathrm{b}$ and the known vacuum
expectation  value (VEV) $v$ of the SM Higgs field,  $\lambda^{\rm
SM}_\mathrm{b} = m_\mathrm{b}/v$. 

In contrast, in the MSSM, bottom-associated production  of neutral
Higgs bosons, $\phi^0 = \mathrm{h}^0$, $\mathrm{H}^0$,
$\mathrm{A}^0$, can dominate the total cross section at large
$\tan\beta$. Two different Higgs doublets are needed to generate
masses  for up- and down-type fermions. These two Higgs doublets
$\mathrm{H_u}$  and $\mathrm{H_d}$  acquire VEVs $v_\mathrm{u}$ and
$v_\mathrm{d}$, respectively, and one defines
$\tan\beta=v_\mathrm{u}/v_\mathrm{d}$. While
$v^2=v_\mathrm{u}^2+v_\mathrm{d}^2$ is fixed by the gauge-boson
masses, the ratio $\tan\beta$ is a free parameter. For large
$\tan\beta$ the down-type  VEV $v_\mathrm{d}$ is small and the
Yukawa coupling of the down-type Higgs doublet is enhanced with
respect to its SM value. The mass $m_\mathrm{b}$  is not small due
to a small Yukawa coupling, on the  contrary, the relevant VEV
$v_\mathrm{d}$ is small. For $\tan\beta \sim \mathcal{O}(50)$, the
bottom Yukawa coupling is  as big as the top Yukawa coupling in the
SM.

The couplings of the CP-even Higgs-boson mass eigenstates are
determined by the mixing of up- and down-type Higgs  fields
characterized by the mixing angle $\alpha$.  For the Yukawa
couplings to b quarks one finds in the MSSM 
\begin{equation}
\label{eq:mssmyukawa}
\begin{array}{ccccc}
\lambda_\mathrm{b}^{\mathrm{h}^0} & = & 
       - \lambda^{\rm SM}_\mathrm{b} 
         \,\frac{\displaystyle \sin\alpha}{\displaystyle \cos\beta} 
	 \, , \\[2.5mm]
\lambda_\mathrm{b}^{\mathrm{H}^0} & = & 
         \lambda^{\rm SM}_\mathrm{b}
	 \,\frac{\displaystyle \cos\alpha}{\displaystyle \cos\beta} 
	 \, , \\[2.5mm]
\lambda_\mathrm{b}^{\mathrm{A}^0} & = & 
       - \lambda^{\rm SM}_\mathrm{b} 
         \, \tan\beta \, .
\end{array}
\end{equation} 
For sizeable mixing in the Higgs sector, the cross sections
$\sigma$  for the b-associated production of all neutral MSSM Higgs
bosons  are enhanced for large $\tan\beta$, i.e.\  $\sigma \propto
\tan^2\beta$. However,  one finds $\sin\alpha \to -\cos\beta$ if the
mass $M_\mathrm{A}^0$ of the  pseudoscalar Higgs boson, the second
input parameter of the MSSM Higgs sector, is large compared to the
Z-boson mass. In this limit, the lighter CP-even Higgs boson
$\mathrm{h}^0$ is SM like and shows no enhanced bottom  Yukawa
coupling. Nevertheless, the total cross section for the two heavy
neutral Higgs bosons $\mathrm{H}^0$ and $\mathrm{A}^0$ is dominated
by b-associated production.

Current searches for bottom--Higgs associated production in the 
MSSM at the Fermilab Tevatron exclude values  $\tan\beta
\;\raisebox{-.3em}{$\stackrel{\displaystyle >}{\sim}$}\; 50$ for
light  $M_\mathrm{A}^0 \approx
100\unskip\,\mathrm{GeV}$~\cite{Abulencia:2005kq}. For a recent
sensitivity study at the LHC see \mbox{Ref.~\cite{Gennai:2007ys}}.

The theoretical description for 
$\mathrm{b}\bar{\mathrm{b}}\,\phi^0$ production can start from
different  initial states for the hard scattering process.  The b
quarks are either generated from gluon splittings within  the hard
process or they are considered to be part of the proton, i.e.\ the
gluon splitting is factorized from the hard process.

In the so-called four-flavour number scheme (4FNS) with no b  quarks
in the initial state, the lowest-order QCD production processes  are
gluon--gluon fusion and quark--antiquark annihilation, 
$\mathrm{g}\mathrm{g} \to \mathrm{b}\bar \mathrm{b}\,\phi^0$ and 
$\mathrm{q}\bar \mathrm{q} \to \mathrm{b}\bar \mathrm{b}\,\phi^0$,
respectively. In this framework the splitting of gluons into
$\mathrm{b}\bar \mathrm{b}$ pairs is treated retaining the full
dependence on the bottom mass. The complete kinematics of the $2 \to
3$ process is available so that the bottom jets in the final state
can be used for tagging and background suppression.

However, for a hard process involving a large scale, e.g.\ the 
Higgs-boson mass, the b quark is effectively almost  massless. Higgs
production is thus dominated by events where gluons split into
nearly collinear $\mathrm{b}\bar \mathrm{b}$ pairs. Consequently,
the inclusive cross section for $\mathrm{g}\mathrm{g} \to
\mathrm{b}\bar \mathrm{b}\,\phi^0$ contains large logarithms
$\ln(\mu_F/m_\mathrm{b})$, where the large scale $\mu_F \sim
M_{\phi^0}$ corresponds to the upper limit of the collinear region
up to which factorization is valid. Hence, the perturbative 
expansion in $\ln(\mu_F/m_\mathrm{b}) \alpha_{\mathrm{s}}$ will
eventually  break down for large Higgs masses and the perturbation
series has to be reorganized. The logarithms
$\ln(\mu_F/m_\mathrm{b})$ can be summed to all orders in
perturbation theory by introducing bottom parton densities. This
defines the so-called five-flavour number scheme
(5FNS)~\cite{Barnett:1987jw}. In this scheme, the leading-order (LO)
process for the inclusive $\mathrm{b} \bar \mathrm{b}\,\phi^0$ cross
section is $\mathrm{b}\bar \mathrm{b}$ fusion,
\begin{equation}
\mathrm{b}\bar \mathrm{b}\to \phi^0\,.
\label{eq:procs_bbfus}
\end{equation}
The next-to-leading order (NLO) cross section in the 5FNS includes
${\cal O}(\alpha_{\mathrm{s}})$ corrections to $\mathrm{b}\bar
\mathrm{b}\to \phi^0$ and tree-level processes like
$\mathrm{g}\mathrm{b}\to\mathrm{b}\phi^0$.  For developments on 
corrections to the latter subprocess see
\mbox{Ref.~\cite{Dawson:2007ur}}.

The use of bottom distribution functions is based on the collinear 
approximation, i.e.\ outgoing b quarks are considered to have small
transverse momentum and to be part of the  proton remnant. There is
no theoretical control over additional b jets at LO.

To all orders in perturbation theory the four- and five-flavour
schemes are identical, but the way of ordering the perturbative
expansion is different, and the results do not match exactly at
finite order. However, numerical comparisons between calculations of
inclusive Higgs production in the two schemes
\cite{Dittmaier:2003ej,Campbell:2004pu,Dawson:2005vi,Buttar:2006zd}
show that the two approaches agree within their respective
uncertainties, once higher-order QCD corrections are taken into
account. 

There has been considerable progress in improving the cross-section
predictions for inclusive associated $\mathrm{b}\bar
\mathrm{b}\,\phi^0$ production by calculating
NLO-QCD~\cite{Dittmaier:2003ej,Dawson:2005vi} and
SUSY-QCD~\cite{Hollik:2006vn} corrections in the four-flavour
scheme, and NNLO QCD corrections~\cite{Dicus:1998hs} in the
five-flavour scheme. In the 5FNS, the QCD scale uncertainties have 
been reduced to the 10\% level such that radiative effects from the 
electroweak sector become of interest.

The complete one-loop QCD and electroweak corrections for the decay
of MSSM Higgs bosons to bottom quarks have been presented in
Ref.~\cite{Dabelstein:1995js}. Recently, complete supersymmetric QCD
and electroweak corrections to the  hadronic production cross
section have been presented  in \mbox{Ref.~\cite{Dittmaier:2006cz}}.
These results and their relation to known  universal corrections are
discussed in the following sections. For technical details and 
derivations we refer the reader to
\mbox{Ref.~\cite{Dittmaier:2006cz}}.

\section{Radiative corrections}
\label{se:rad}

In b-quark fusion, $\mathrm{b}\bar \mathrm{b}\to \phi^0$, there are
universal radiative  corrections which lead to the definition of the
improved Born approximation for the partonic cross section
\begin{equation}
\label{eq:IBA}
\hat{\sigma}_{\rm {IBA}} \, = \, 
\hat{\sigma}_{\rm {SM}}
\left\{ 
\begin{array}{l} 
\frac{\displaystyle \sin^2 {\alpha_{\rm eff}}}
     {\displaystyle\cos^2\beta}
\left(\frac{\displaystyle 1 - \Delta_\mathrm{b}/
            (\tan\beta \tan {\alpha_{\rm eff}})}
	   {\displaystyle 1 + \Delta_\mathrm{b}}
	   \right)^2 \,  \\[2ex]
\frac{\displaystyle \cos^2 {\alpha_{\rm eff}}}
     {\displaystyle\cos^2\beta}
\left(\frac{\displaystyle 1 + \Delta_\mathrm{b} \, 
           \tan {\alpha_{\rm eff}}/\tan\beta}
	   {\displaystyle 1 + \Delta_\mathrm{b}}
	   \right)^2 \,  \\[2ex]
\,\,\,\,\, \tan^2\beta \,\,\,\, 
\left(\frac{\displaystyle 1 - 
            \Delta_\mathrm{b}/\tan^2\beta}
	   {\displaystyle 1 + \Delta_\mathrm{b}}
	   \right)^2 \, ,
\end{array} 
\right. 
\end{equation}
where
\begin{equation}
\hat{\sigma}_{\rm {SM}} \, =\, 
\frac{\sqrt{2}\pi G_\mu\overline{m}_{\mathrm{b}}(\mu_R)^2}
     {6 M_{\phi^0}^2} \, \, \delta(1-M_{\phi^0}^2/\hat{s}) \, .
\end{equation}
We denote the partonic CMS energy by $\sqrt{\hat{s}}$ , 
$M_{\phi^0}$ is the mass of the produced Higgs boson,  $G_\mu$ is
the Fermi constant, and $\overline{m}_{\mathrm{b}}(\mu_R)$ is the
running bottom mass at the renormalization scale $\mu_R$. 
Potentially large radiative  corrections are encoded in the
parameters $\Delta_\mathrm{b}$ and $\alpha_{\rm eff}$ to be  briefly
explained in the following. 

While b quarks do not couple to the Higgs field $\mathrm{H_u}$ at 
tree level, this interaction is radiatively induced at the  one-loop
level, e.g.\ via the sbottom coupling to $\mathrm{H_u}$ in SUSY-QCD
loops. This induces a shift $\Delta_\mathrm{b}$ in the relation
between the b mass and the respective Yukawa coupling. This shift
is  proportional to $\tan\beta$ and, thus, for large $\tan\beta$
the  corresponding correction is sizeable. It has been 
shown~\cite{Carena:1999py} that the correction can be resummed and
that it affects the cross section  according to eq. (\ref{eq:IBA}).

\begin{table*}[t]
\begin{center}
\begin{tabular}{|l|c|c|c|c|c|c|}
\hline
& \multicolumn{2}{|c|}{\raisebox{-1pt}{$\mathrm{h}^0$}} & 
\multicolumn{2}{|c|}{\raisebox{-1pt}{$\mathrm{H}^0$}} &
\multicolumn{2}{|c|}{\raisebox{-1pt}{$\mathrm{A}^0$}} \\ \cline{2-7}
&  $m_\mathrm{b}$[GeV]  & $\sigma$[pb] & 
   $m_\mathrm{b}$[GeV]  & $\sigma$[pb] & 
   $m_\mathrm{b}$[GeV]  & $\sigma$[pb] \\ \hline
  QCD                                  &  2.80 &     0.97 &     2.55 &    24.12 &     2.55 &    24.13 \\ 
 +QED                                  &  2.80 &     0.97 &     2.55 &    24.07 &     2.55 &    24.09 \\ 
$+\Delta_{\mathrm{b}}^{\tilde{g}}$     &  2.72 &     0.92 &     1.95 &    14.14 &     1.95 &    14.15 \\ 
$+\Delta_{\mathrm{b}}^{\mathrm{weak}}$ &  2.75 &     0.94 &     2.24 &    18.66 &     2.24 &    18.67 \\ 
$+ \sin (\alpha_{eff})$                &  2.75 &     0.88 &     2.24 &    18.66 &     2.24 &    18.67 \\ \hline
  full calculation                     &  2.75 &     0.87 &     2.24 &    18.43 &     2.24 &    18.44 \\ 
\hline
\end{tabular}
\end{center}
\caption{\label{Tab:results_SPS4_cross_sections} 
  The  effective bottom mass and the NLO MSSM cross section
  $\mathrm{p}\mathrm{p}\to
  (\mathrm{b}\bar{\mathrm{b}})\,
  \mathrm{h}^0/\mathrm{H}^0/\mathrm{A}^0\!+\!X$
  at the LHC ($\sqrt{s}=14$~TeV) in the SPS~4 scenario including the
  cumulative corrections due to the different classes of
  corrections. See text for details on the different contributions. 
  (Table taken from \mbox{Ref.~\cite{Dittmaier:2006cz}}) 
}
\end{table*}

Radiative corrections can also have a large impact on the  mixing of
the Higgs fields to form the CP-even mass eigenstates $\mathrm{h}^0$
and $\mathrm{H}^0$. The bulk of these corrections can be absorbed
in  a loop-corrected, process-independent effective mixing angle 
$\alpha_\mathrm{eff}$ which replaces its tree-level counterpart in
eq. (\ref{eq:IBA}). 

Precise definitions of $\Delta_\mathrm{b}$ as well as
$\alpha_\mathrm{eff}$ are given in
\mbox{Ref.~\cite{Dittmaier:2006cz}}, where we also describe in
detail the calculation of the complete SUSY-QCD  and electroweak
corrections including the renormalization of the  MSSM Higgs sector.
The bottom mass has been renormalized  in such a way that the
corrections due to $\Delta_\mathrm{b}$ are absorbed  into the input
value for $m_\mathrm{b}$. Thus, the  numerical value for this
effective $m_\mathrm{b}$ quantifies the  $\tan\beta$ enhanced
corrections. This procedure automatically avoids double counting for
the $\Delta_\mathrm{b}$ corrections.  When we relate the 
corrections from the full calculation to $\hat{\sigma}_{\rm {IBA}}$
we also carefully avoid double counting with respect to corrections
already contained in $\alpha_\mathrm{eff}$.

\section{Results}
\label{se:MSSM_results}

All the results in this section are calculated in the 
$\overline{\mathrm{DR}}$ scheme for $\tan\beta$. The 
renormalization and factorization scales are set to $\mu_R =
M_{\phi^0}$ and $\mu_F = M_{\phi^0}/4$,  respectively. We use the 
MRSTQED2004 PDF\cite{Martin:2004dh} which also allows the inclusion
of the photon-induced partonic channels at NLO. The input-parameter
scheme and the numerical input are specified in
\mbox{Ref.~\cite{Dittmaier:2006cz}}. To further improve our NLO
results, we use two-loop improved Higgs self-energies provided by
the program package  {\tt FeynHiggs}~\cite{FeynHiggs}.

Within the MSSM, we first focus on the radiative corrections and
total cross sections in the SPS~4 benchmark scenario  ($\tan\beta =
50$)~\cite{Allanach:2002nj} which was designed to give large cross
sections for heavy Higgs bosons. At the end of this section we also
show results for the other SPS points. While most of the SPS
scenarios are in conflict with experimental data by now, they are
still valuable because they cover typical SUSY scenarios in
different regions of parameter space.

In Table~\ref{Tab:results_SPS4_cross_sections}, we show the effect
of the various higher-order corrections on the effective 
$\mathrm{b}$-mass. Starting from the running QCD mass at the  scale
of the Higgs-boson mass, the shifts from SUSY-QCD 
($\Delta_{\mathrm{b}}^{\tilde{g}}$) and the electroweak sector 
($\Delta_{\mathrm{b}}^{\mathrm{weak}}$) are included.  The
corresponding cross sections $\sigma$ are first shown at NLO QCD. 
As can be seen, the QED corrections are generally very small after
mass factorization. The summation of the $\tan\beta$-enhanced
MSSM-QCD and MSSM-weak corrections has a significant effect on the
cross sections for $\mathrm{H}^0$ and $\mathrm{A}^0$ production. The
light Higgs boson $\mathrm{h}^0$ is SM-like and the summation of
terms $\propto \tan\beta$ has thus no sizeable impact. Employing a
loop-improved effective mixing angle $\alpha_{\rm eff}$ is
numerically relevant only for $\mathrm{h}^0$ production. The cross
sections in the last-but-one row of 
Table~\ref{Tab:results_SPS4_cross_sections} correspond to the
improved Born approximation $\sigma_{\rm {IBA}}$  dressed with QCD
and QED corrections. The full MSSM cross sections, including all
summations and the remaining non-universal ${\cal
O}(\alpha_{\mathrm{s}})$ and ${\cal O}(\alpha)$ corrections, are
displayed in the last row of the table. 

The bulk of the MSSM-QCD and -weak corrections can indeed be
absorbed into the above definition of $\sigma_{\rm {IBA}}$. The
remaining non-universal corrections in the complete MSSM calculation
turn out to be quite small, below approximately 2\%. 

\begin{figure}[b]
\begin{center}
\begin{picture}(0,0)(0,0)
\put(110,-8){$M_\mathrm{A}^0$[GeV]}
\put(0,170){$\delta_{\mathrm{MSSM}}$[\%]}
\end{picture}
\includegraphics[width=.48\textwidth]{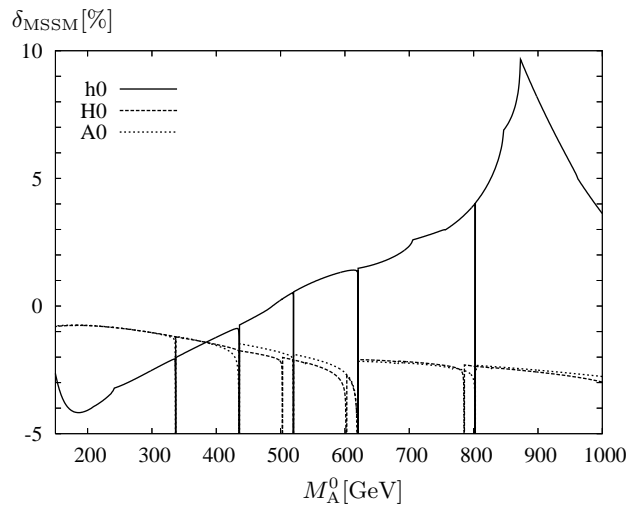}
\end{center}
\caption{\label{Fig:MA_plot} 
  Full MSSM corrections $\delta_{\rm  MSSM}$
  defined relative to $\sigma_{\mathrm{IBA}}$ as a function of the
  $M_\mathrm{A}^0$ pole mass. All other MSSM parameters are fixed 
  to their SPS~4 values. 
  (Figure taken from \mbox{Ref.~\cite{Dittmaier:2006cz}})}
\end{figure}
\begin{figure}[t]
\begin{center}
\begin{picture}(0,0)(0,0)
\put(110,-8){$m_\mathrm{b}$[GeV]}
\put(0,170){$\delta_{\mathrm{MSSM}}$[\%]}
\end{picture}
\includegraphics[width=0.48\textwidth]{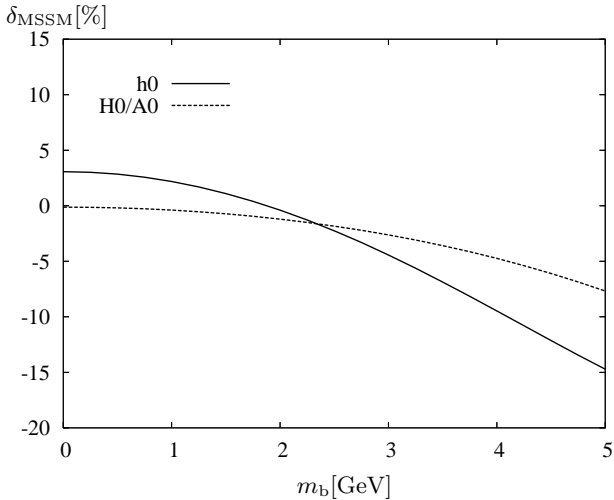}
\end{center}
\caption{\label{Fig:MB_plot} 
  Full MSSM corrections $\delta_{\rm MSSM}$
  defined relative to $\sigma_{\mathrm{IBA}}$ as a function of the
  $m_\mathrm{b}$ input. The corrections for $\mathrm{H}^0$ and 
  $\mathrm{A}^0$ lie on top of each other. 
  (Figure taken from \mbox{Ref.~\cite{Dittmaier:2006cz}})}
\end{figure}

In Fig.~\ref{Fig:MA_plot} we show the impact of the complete
supersymmetric ${\cal O}(\alpha_{\mathrm{s}})$ and ${\cal
O}(\alpha)$ corrections defined relative to the improved Born
approximation $\sigma_{\rm IBA}$ for different values of the
on-shell mass $M_\mathrm{A}^0$.   All other MSSM parameters are kept
fixed at their SPS~4 values. The size of the non-universal
corrections does not exceed $3\%$ for $\mathrm{H}^0/\mathrm{A}^0$
production except for special model parameters where the Higgs
masses are close to the production threshold for pairs of
sparticles.  These unphysical singularities can be removed by taking
into account  the finite widths of the unstable sparticles.  The
size of $\delta_{\mathrm{MSSM}}$ for $\mathrm{h}^0$ depends very
sensitively on the definition of the effective mixing angle
$\alpha_{\rm eff}$ employed in $\sigma_{\rm IBA}$. Note that in any 
case $\mathrm{h}^0$ is SM-like at large $M_\mathrm{A}^0$ so that 
$\mathrm{h}^0$ production in bottom fusion is most likely of no
phenomenological relevance.

It is important to emphasize that the non-universal MSSM corrections
$\delta_{\mathrm{MSSM}}$ at large $\tan\beta$ are quite sensitive to
the choice of the b-mass input value within the one-loop corrections
which is not fixed by the renormalization procedure.  There are
terms that grow as $m_\mathrm{b}^2 \tan^2\beta$ which are not
included in $\Delta_\mathrm{b}$.  For the SPS~4 scenario the
sensitivity on the $\mathrm{b}$-mass input is shown in
Fig.~\ref{Fig:MB_plot}. The absolute size of the non-universal
corrections varies between approximately zero and $-6\%$  for the
phenomenologically relevant  $\mathrm{H}^0/\mathrm{A}^0$ production,
depending on whether a massless approximation, the effective running
mass, or the pole mass is chosen as $\mathrm{b}$-mass input.
Although we assume that the running mass, including the corrections
from $\Delta_\mathrm{b}$ (as used for all the shown results), is a
sensible choice, the sensitivity of the NLO correction to the
$\mathrm{b}$-mass input constitutes a theoretical uncertainty which
cannot be resolved at the NLO level.

Table~\ref{ta:sps} displays the cross sections along with the 
non-universal corrections from the full calculation for the
different  SPS points. It shows that these residual corrections are
small and do not exceed 2\% for $\mathrm{H}^0/\mathrm{A}^0$
production in a wide range of MSSM parameters.

\begin{table}[b]
\begin{center}
\begin{tabular}{|l|r|r|r|r|r|r|}
\hline
       &  \multicolumn{3}{|c|}{$\sigma [\mathrm{pb}]$}   &
          \multicolumn{3}{|c|}{$\delta [\%]$}             \\ \hline
\raisebox{-1pt}{SPS}    
       &  \multicolumn{1}{|c|}{\raisebox{-1pt}{$h^0$}}
       &  \multicolumn{1}{|c|}{\raisebox{-1pt}{$H^0$}}
       &  \multicolumn{1}{|c|}{\raisebox{-1pt}{$A^0$}}  
       &  \multicolumn{1}{|c|}{\raisebox{-1pt}{$h^0$}}  
       &  \multicolumn{1}{|c|}{\raisebox{-1pt}{$H^0$}}  
       &  \multicolumn{1}{|c|}{\raisebox{-1pt}{$A^0$}}  \\ \hline
 1a &$   1.03 $&$    0.91 $&$	 0.92 $&$    2.29 $&$	-0.21 $&$    0.15 $\\
 1b &$   0.81 $&$    2.23 $&$	 2.23 $&$    1.96 $&$	-0.20 $&$   -0.21 $\\
 2  &$   0.77 $&$    0.00 $&$	 0.00 $&$    3.11 $&$	-1.35 $&$   -1.35 $\\
 3  &$   0.84 $&$    0.18 $&$	 0.18 $&$    4.17 $&$	 0.02 $&$    0.00 $\\
 4  &$   0.87 $&$   18.43 $&$	18.44 $&$   -0.92 $&$	-1.24 $&$   -1.27 $\\
 5  &$   0.95 $&$    0.02 $&$	 0.02 $&$   -4.08 $&$	 0.26 $&$   -1.10 $\\
 6  &$   0.95 $&$    0.47 $&$	 0.47 $&$    3.06 $&$	-0.12 $&$    0.19 $\\
 7  &$   1.09 $&$    2.45 $&$	 2.46 $&$    4.62 $&$	 1.59 $&$    1.61 $\\
 8  &$   0.92 $&$    0.67 $&$	 0.67 $&$    5.86 $&$	 0.96 $&$    1.25 $\\
 9  &$   0.83 $&$    0.02 $&$	 0.02 $&$    3.36 $&$	-0.87 $&$   -0.81 $\\ \hline
\end{tabular}
\end{center}
\caption{
Cross sections $\sigma$ and non-universal corrections $\delta$
for Higgs production in the SPS scenarios. $\delta$ is given with 
respect to $\sigma_\mathrm{IBA}$ being dressed with NLO QCD/QED 
corrections.
}
\label{ta:sps}
\end{table}

\section{Conclusions}
\label{se:conclusion}

We have given a brief review on Higgs production in association 
with bottom quarks focussing on the basic concepts and precise
predictions for Higgs production in bottom-quark fusion. The leading
supersymmetric higher-order corrections can be taken  into account
by an appropriate definition of the couplings in an improved Born
approximation. The remaining non-universal corrections are small,
typically of the order of a few per cent. The theoretical
uncertainty connected to the input value of the b-quark mass within
the one-loop correction is emphasized.

\section*{Acknowledgments}

We are grateful to Stefan Dittmaier, Michael Kr\"amer, and Tobias 
Schl\"uter for the collaboration on the original 
work~\cite{Dittmaier:2006cz} and to Ansgar Denner, Stefan
Ditt\-maier,  and Michael Kr\"amer for comments on the manuscript.

\end{document}